\documentclass{iopart}

  \expandafter\let\csname equation*\endcsname\relax
  \expandafter\let\csname endequation*\endcsname\relax

\usepackage[english]{babel}
\usepackage[T1]{fontenc}
\usepackage{lmodern}
\usepackage{dsfont}

\usepackage[utf8]{inputenc}
\usepackage{uniinput}

\usepackage{graphicx}
\graphicspath{{figs/}}

\usepackage[sort&compress,numbers]{natbib}

\begin{document}

\title{Efficient simulation of non-Markovian system-environment
interaction}

\author{Robert Rosenbach${^1}$, Javier Cerrillo$^{2,3}$$^*$, Susana F.
Huelga$^{1}$, Jianshu Cao$^{2}$, Martin B. Plenio$^{1}$$^*$}

\address{
$^1$ Institute of Theoretical Physics, Universit{\"a}t Ulm,
Albert-Einstein-Allee 11, 89069 Ulm, Germany}
\address{
$^2$ Massachusetts Institute of Technology, 77 Massachusetts Avenue,
Cambridge, Massachusetts 02139, USA}
\address{
$^3$ Institute of Theoretical Physics, Technische Universität Berlin,
Hardenbergstr. 36, 10623 Berlin, Germany}
\address{$^*$ Corresponding authors: cerrillo@tu-berlin.de, martin.plenio@uni-ulm.de}

\begin{abstract}

\noindent In this work, we combine an established method for open quantum
systems -- the time evolving density matrix using orthogonal polynomials
algorithm (TEDOPA) -- with the transfer tensors formalism (TTM), a new tool for the
analysis, compression and propagation of non-Markovian processes. A compact propagator is generated out of sample trajectories covering the correlation time of the bath.
This enables the investigation of previously inaccessible long-time
dynamics with linear effort, such as those ensuing from low temperature regimes with arbitrary, possibly highly structured,
spectral densities. We briefly introduce both methods,
followed by a benchmark to prove viability and combination synergies.
Subsequently we illustrate the capabilities of this approach at the hand of specific examples and
conclude our analysis by highlighting possible further applications of
our method.

\end{abstract}

\maketitle

\section{Introduction}

Ranging from condensed matter physics or quantum technologies to
biological chemistry, the experimental ability to accurately probe and
analyse quantum systems in strong contact with highly structured
environmental degrees of freedom for extended periods of time has
become a reality \cite{ref:qdots,ref:spins2015,ref:NV,ref:Engel2007,ref:Panit2010,
ref:Collini2010}. Under certain conditions it is possible to model the
observations by using approximate methods such as perturbative
approaches \cite{ref:breuer2007, ref:rivas2011}, frequently supplemented with Markovian assumptions
\cite{ref:Linblad, ref:gorini1976,ref:Redfield}.
Beyond their regime of validity, the task of exactly treating the
dynamics of open quantum systems faces the challenge of an unfavorable
scaling in the required resources.  Nevertheless, many tools have been
developed that address a wide variety non-Markovian scenarios for short
time simulations or for specific conditions and approximations.

Exact procedures such as projection operator techniques serve to derive
formally exact master equations that can involve a memory kernel as in
the Nakajima-Zwanzig formalism \cite{ref:Nakajima} or have a generator
that is local in time, like the so-called time-convolutionless master
equations \cite{ref:bassano}. The path integral formulation
\cite{Feynman1948} provides an alternative perspective especially suitable
for harmonic baths thanks to the Feynmann-Vernon influence functional
\cite{Feynman1963}. Practical implementation of these formal treatments
requires perturbative expansions in terms of some specific parameter, be
it weak damping, high temperature, short memory time, or short
simulation time \cite{Cao1997, ref:tanimura1989, Makri1995,
Leggett1987}. An exhaustive list is not within the scope of this work, but some additional instances include the non-Markovian quantum state diffusion approach \cite{ref:NMQSD}, the hierarchical equations of motion (HEOM) \cite{ref:tanimura1989,Tang2015}, iterative path integral resummations \cite{Weiss2008}, multilayer multiconfigurational time-dependent Hartree methods  \cite{Manthe2008}, explicit computation of the Nakajima-Zwanzig memory kernel \cite{Shi2003,Cohen2011} and the time-convolutionless kernel \cite{Kidon2015} or mixed quantum-classical methods \cite{Kapral2015,Kelly2013,Gottwald2015}.
 Alternatively, the harmonic bath assumption renders possible the
use of stochastic Gaussian sampling of the bath operator or the
influence functional \cite{Cao1996, Egger2000, ref:Stockburger2002, Muhlbacher2008}. Hybrid stochastic-deterministic methods have appeared recently as well \cite{ref:moix2013}. Another option is to simulate the density matrix of both the system and
the complete environment by employing an efficient description of the
bath or gradually introducing select degrees of freedom \cite{Baer1997,Gualdi2013,DeVega2014}. To this class of methods belongs the ``time evolving density
matrix using orthogonal polynomials algorithm'' (TEDOPA)
\cite{ref:prior2010, ref:chin2010}. This method uses a stable numerical transformation to
map the environment into a chain of harmonic oscillators, which can then be simulated together with the system using efficient quantum
many body techniques \cite{ref:dmrg}. Although a more thorough discussion follows below, as compared to other simulation methods TEDOPA is especially suitable for simulation of quadratic harmonic baths with arbitrarily shaped spectral-densities in the low-temperature regime and is not restricted to small coupling or Ohmic baths.

For any exact simulation method it is generally the case that the size
of the propagator or that of the stochastic sample scales unfavorably with
the time length of the simulation or the corresponding perturbative
expansion order. Then the question arises whether there are regimes where
this scaling can be mitigated in some form, i.e.~if an effective
propagator of a reduced size can be extracted with the intention of
facilitating long-time simulations. In the present work we address this
question combining TEDOPA with a tool quantifying the bath's
back-action on the system as in the Nakajima-Zwanzig formalism. This
tool is known as the transfer tensor method (TTM)
\cite{ref:cerrillo2014}, which has been shown to provide considerable
acceleration in the context of non-Markovian open quantum system
simulations as well as in large classical systems
\cite{ref:mehraiid2015}. This is achieved by blackbox learning from
sample exact trajectories for some short initial period and subsequent
generation of a compact multiplicative propagator for the system
degrees of freedom alone.
This propagator is the set of discrete elements of the integration of the Nakajima-Zwanzig equation which, similarly to the memory kernel, decay at the rate of the bath correlation function. This justifies the definition of a memory cutoff, corresponding to the maximum time for which non-Markovian effects are to be considered.  The method does not require input of a microscopic description of the problem and just involves straightforward analysis of the evolution of the system density matrix. In this sense, TTM is directly applicable to any state propagation simulator. For a learning period longer than the
environment correlation time, the propagator accurately reproduces the
long time system dynamics with linear effort. Another possibility to exploit the decay of the memory effects in dissipative systems is to explicitly calculate the Nakajima-Zwanzig memory kernel \cite{Shi2003,Cohen2011}. Although this is in general a demanding task, it is possible for specific systems and has been implemented with the help of quantum Monte-Carlo methods \cite{Cohen2011,Cohen2013} or multilayer multiconfigurational time-dependent Hartree methods \cite{Wilner2013,Wilner2014,Wilnerb2014,Thoss}.
With TTM an explicit computation of the memory kernel is avoided and a discretized propagator is directly obtained. This propagator grows linearly with the correlation time of the bath, improving on the size of some deterministic simulation methods of linear propagation effort \cite{ref:tanimura1989, Makri1995}. It is a general and
flexible approach that does not depend on the form of the environment
or the interaction, while TEDOPA is not restricted to weak system-bath
coupling, high temperatures or specific spectral densities. Therefore,
they constitute ideal partners and a study of their combined
performance represents a natural research question.

We start the discussion in section \ref{sec:Method} by providing a general analysis
of both TEDOPA and TTM, thereby specifying the regime in which their
combination is expected to be most productive. In addition, tools for
the error assessment are provided. In section \ref{sec:Benchmark} we provide a
benchmark between the proposed combination and the exact result and confirm perfect agreement. Finally, in section \ref{sec:Applications} relevant applications
of the TEDOPA-TTM combination are proposed which include Ohmic and
non-Ohmic spectral densities, low temperature simulations and computations of
absorption spectra.

\section{The Method \label{sec:Method}}

\subsection{A numerically exact open quantum system simulator}

TEDOPA \cite{ref:prior2010, ref:chin2010} is a numerically exact and
certifiable simulation method for open quantum system dynamics
\cite{ref:woods2015}. It applies to general systems under linear
interaction with an environment modeled by a set of independent
harmonic oscillators such that the total Hamiltonian $H$ can be split
into the system, the environment and the interaction between the two as
\begin{align}
	&H = H_{\text{sys}} + H_{\text{env}} + H_{\text{int}},
	\label{eq:H_spin_boson}\\
	&H_{\text{env}} = ∫_0^{x_{\text{max}}} dx~g\left( x \right)
			  a_x^† a_x, \label{eq:def_H1} \\
	&H_{\text{int}} = ∫_0^{x_{\text{max}}} dx~h\left( x \right)
			  \left( a_x^† + a_x \right) A .
	\label{eq:def_H2}
\end{align}
Here $a_x^†$ and $a_x$ denote the bosonic creation and annihilation
operators corresponding to the environmental mode $x$. The coefficients
$g\left(x \right)$ can be identified as the environmental dispersion
relation. The interaction term $H_{\text{int}}$ assigns each mode a
coupling strength~$h\left(x\right)$ between its displacement $a_x^† +
a_x$ and a general system operator $A$.

Together with the temperature, the functions $g\left(x\right)$ and
$h\left( x \right)$ characterize the harmonic environment uniquely and
define the {\em spectral density} $J\left( ω \right)$ as
\begin{equation}
	J\left( ω \right) = π h²\left[ g^{-1}\left( ω \right) \right]
	\frac{dg^{-1}\left( ω \right)}{dω}.
	\label{eq:def-sd}
\end{equation}
Here $g^{-1}\left[ g\left( x \right) \right] = g\left[ g^{-1}\left( x
\right) \right] = x$ and $g(x)$ is monotonically growing. The quantity $\frac{dg^{-1}\left( ω
\right)}{dω}δω$ can be interpreted as the number of quantized modes
with frequencies between ω and ω+δω (for δω→0). We consider spectral
densities subject to a hard cut-off at frequency~$ω_{\text{hc}}$; this
in turn defines the cut-off $x_{\text{max}}$ in Eqs.~\eqref{eq:def_H1}
and \eqref{eq:def_H2} as $x_{\text{max}}=g^{-1}(ω_{\text{hc}})$.

TEDOPA uses a two-stage sequence to enable full treatment of the system
and environment degrees of freedom. In a first step, an analytic
transformation based on orthogonal polynomials converts the
star-shaped system-environment structure into a one-dimensional geometry, where the
system couples only to the first site of a semi-infinite chain of harmonic oscillators that
contains only nearest-neighbour interactions. This is accomplished by
use of a unitary transformation $U$, which defines new harmonic
oscillators with creation and annihilation operators $b_n^†$, $b_n$ given by
\begin{align}
	&U_n\left( x \right) = h\left( x \right) p_n\left( x \right),\\
	&b_n^† = ∫_0^{x_{\text{max}}} dx~U_n\left( x \right) a_x^†.
	\label{eq:chainmap}
\end{align}
Here $p_n\left( x \right)$ are orthogonal polynomials defined with
respect to the measure $dμ\left( x \right)=h²\left( x \right)dx$ and the three-term recurrence relation
\begin{equation}
	p_{k+1}\left( x \right) =
	\left( x - α_k \right) p_k\left( x \right) -
	β_k p_{k-1}\left( x \right),
	\label{eq:defmonicrecurrel}
\end{equation}
with $p_{-1}(x)\equiv0$ and $k$ a positive integer or zero. This
transformation can be performed analytically for specific spectral
densities \cite{ref:burkey1984, ref:chin2010}. For arbitrarily shaped
spectral densities a numerically stable approach has been developed
\cite{ref:prior2010, ref:gautschi1994}. Mappings using orthogonal polynomials have been shown to be exact for quadratic Hamiltonians \cite{ref:DeVega2015}.

The recurrence relation (\ref{eq:defmonicrecurrel}) results in the one-dimensional
configuration with the Hamiltonian
\begin{align}
	\tilde{H} &= H_{\text{sys}}
	  + η_0 A \left( b_0 + b_0^† \right)
	  + \sum_{n=0}^∞ ω_n b_n^† b_n \nonumber \\
	  &+ \sum_{n=0}^∞ η_n \left( b_n^† b_{n+1} + b_n b_{n+1}^† \right).
	\label{eq:defchainh}
\end{align}
For a linear dispersion relation $g(x)=g' x$, $\omega_n = g' \alpha_n $ and $\eta_n = g' \sqrt{\beta_{n+1}}$.
Due to the form of the emerging linear geometry, this first stage of TEDOPA is
coined ``chain mapping''. One dimensional quantum many body systems can
be efficiently simulated with the well established time dependent
density matrix renormalization group (t-DMRG) algorithm
\cite{ref:schollwoeck2011,ref:vidal2004,ref:White2004}. Its application to evaluate
the dynamics of the system {\em and} the transformed environment
constitutes the second stage of TEDOPA.  The long ranged correlations
appearing in the original star-shaped geometry advise against the
application of t-DMRG in that picture: it is the nearest neighbor
structure that makes the numerical t-DMRG approach particularly
efficient. Recent works consider the possibility to use generalized matrix product state formulations for treatment of star-shaped baths as well \cite{Wolf2014}.

For a complete account of TEDOPA's inner workings refer to
\cite{ref:prior2010, ref:chin2010, ref:woods2014}. Suffice it to say
that three main aspects distinguish it from other open quantum-system
methods. First, it does not restrict the ratio~$λ$ between inner-system
couplings and system-environment couplings, unlike numerous other
methods (which assume either $λ\gg1$ or $λ\ll1$).  Second, the spectral
density characterizing the system-environment interaction may assume
any arbitrary shape, including a wide variety of sharp features that
may be related to long-lived vibrational modes \cite{ref:prior2010,ref:chin2013, ref:dijkstra2015}.
Such spectral densities are often encountered in spectral densities
reconstructed from experimental data, e.g. in biological settings
\cite{ref:adolphs2006}.  And third, while applicable to all
temperatures, due to its scaling properties TEDOPA is inherently
well-suited for simulations in the low-temperature domain.

Naturally, however, exact numerical methods tend to be costly and an
effort to save on the associated computational demands is desirable.
Where the transfer tensor method can be applied, these costs can be
reduced and challenging long time simulations become accessible.

\subsection{Non-Markovian Dynamical Maps: Transfer Tensor Method}

The transfer tensor method \cite{ref:cerrillo2014} reduces the numerical effort of TEDOPA
simulations for a large class of non-Markovian environments.  Its key
idea is to relate the initial stages of the system's evolution to
later times by efficiently determining the dynamical correlations built
up between system and environment. This is achieved by the reconstruction of
dynamical maps for short initial evolution times and their subsequent transformation into so-called {\em transfer
tensors}.

A dynamical map is defined for an initial condition where the state of the system and the state of
the environment are separable, and it fully determines the reduced state of the system $\rho(t)$ when applied to an initial condition $\rho(0)$
\begin{equation}
	ρ\left( t \right) = \mathcal{E}(t,0) ρ\left( 0 \right).
	\label{eq:def_dynmaps}
\end{equation}
For a time independent Hamiltonian $H$ such as eq.
\ref{eq:H_spin_boson}, eq.\ref{eq:def_dynmaps} is related to the solution of the time-translationally-invariant Nakajima-Zwanzig equation
\begin{equation}
	\dot{ρ}\left( t \right) = -i \mathcal{L}_s ρ + ∫_0^t dt'
	\mathcal{K} \left( t-t' \right) ρ\left( t' \right),
	\label{eq:def_nz}
\end{equation}
where $\mathcal{L}_s\rho=[H_{sys},\rho]$ is the Liouvillian of the
system alone and $\mathcal{K}(t-t')$, is the memory kernel arising from
the system-bath interaction $H_{int}$.

The input for TTM is a set of dynamical
maps~$\mathcal{E}_k=\mathcal{E}(t_k,0)$, where a discretisation time
step $\delta t$ ($t_{k}\equiv k\delta t$) is used.  These dynamical
maps are obtained easily from the evolution of all distinct initial
basis states of the system's density matrix. The transfer tensors are
then iteratively defined by
\begin{equation}
	T_{n} = \mathcal{E}_n - \sum_{m=1}^{n-1} T_{n-m} \mathcal{E}_m.
	\label{eq:T_from_E}
\end{equation}
According to this definition, the transfer tensor $T_k$ then quantifies
the correlation in the dynamical map $\mathcal{E}_k$ with the
non-Markovian effects built up during the previous~$k$ time steps.

Further, the discretization of the memory
kernel~$\mathcal{K}$ is directly related to these tensors by the time
step~$δt$
\begin{equation}
	T_{k} = \mathcal{K}_{k} δt²,
	\label{eq:rel_TK}
\end{equation}
where $\mathcal{K}_{k}=\mathcal{K}\left( t_k \right)$ is the discretized memory
kernel at time $t_k$. The system Hamiltonian
is itself accessible from the first transfer tensor by
\begin{equation}
	T_{1} = (\mathds{1}-i\mathcal{L}_sδt).
	\label{eq:H_from_T}
\end{equation}

These tensors can be used to propagate the system to arbitrary later
times as long as they cover the relevant part of the memory kernel. Assuming a finite coherence time of the bath $t_\text{bath}$, the transfer tensors will decay sufficiently fast that a cutoff $K$ can be defined such that $T_k=0$ for $k>K$. Then, $\rho(t_n)$ for $n>K$ can be expressed simply as
\begin{equation}
\rho(t_n)=\sum\limits_{k=1}^{K}T_k\rho(t_{n-k}).
\label{eq:TTM}
\end{equation}

TTM is applicable in a variety of interesting cases, scaling favorably
in system size and length of the environment's correlation time. Due to
its nature it also does not rely on assumptions about the system's
parameters or environmental couplings. Thus it is usable as a powerful
black box tool which, given initial trajectories, subsequently delivers
the evolution trajectories for later times. For a more complete account
of this tool refer to~\cite{ref:cerrillo2014}.

\subsection{Synergies}
\label{sec:synergies}

The combination of both methods facilitates the simulation of open
quantum systems in regimes that were previously inaccessible.
Exceptionally relevant is the ability to perform long-time simulations
of low-temperature, highly-structured harmonic environments at merely a
fraction of the computational cost which would be necessary if only
TEDOPA were applied. Evidence for this is provided by simple
examination of some of the features of each of the methods. On the one
hand, TEDOPA is based on a matrix product operator (MPO) description of the complete system plus
environment density matrix. Settings leading to low occupation numbers
of environmental oscillators -- such as low temperatures -- are especially
suitable as they reduce the MPO's number and size. In addition TEDOPA
is not limited to a specific analytical form of spectral density. On
the other hand, recurrence effects originating at the chain's boundary
limit the time for which accurate simulation is possible.  Because TTM
only requires sample system trajectories for as long as bath
correlations are present, the required chain length is then not anymore
determined by the target simulation time. The combination of TTM and
TEDOPA is therefore most useful in highly non-Markovian regimes where
bath correlation times are comparable to the maximum simulation time
that TEDOPA can reach before recurrences appear.

\subsection{Parameters and accuracy}

Here we analyse relevant parametric cutoffs for the control of the
accuracy of numerical simulations that combine TEDOPA and TTM.

In the case of TEDOPA one can distinguish between parameters related to
the chain mapping and to the t-DMRG propagation. The semi-infinite
chain of oscillators generated by the mapping necessarily requires the
truncation of both the chain length~$N$ and the Hilbert space dimension~$d$ for each oscillator at a
reasonable value. Those are native TEDOPA error sources and they have a
direct effect on the recurrence time of the simulation and the maximum
temperature that may be simulated, respectively. It should be noted
that the error incurred by these two approximations can be upper-bounded rigorously by analytical expressions \cite{ref:woods2015}. The
effect of other relevant parameters, namely the MPO's matrix
size~$(χ\times\chi)$ and the time step~$δt$, are already well-known
from the time-evolving block decimation (TEBD) algorithm \cite{ref:vidal2004}.  To reach the accuracy required by TTM,
care needs to be taken in adjusting these parameters to bound the total
error of TEDOPA sufficiently. Some indications on how to accomplish
this are provided in the present section.

The maximum time $t_{max}$ before unphysical back-actions of the
environment due to reflections at the end of the chain appear is
related to the chain length $N$. Usually all chain coefficients are of
the same order of magnitude and hence the simulation time~$t_{max}$
scales roughly linearly with~$N$. This reveals one of the benefits of
the application of TTM on TEDOPA: the chain length can be truncated
according to the length of the bath correlation time, allocating the
simulation resources properly and shortening simulation times
considerably in many cases of relevance (cf.
Fig.~\ref{fig:t_simulation}). The exact relationship between $N$ and
$t_{max}$ may be derived analytically through the use of Lieb-Robinson
bounds \cite{ref:woods2015} or numerically by trial-and-error: by
setting the chain in an initial state $|10…0\rangle$ and following the
evolution of the number operator $n$ on the first site,
$O=n\otimes\mathds{1}\otimes…\otimes\mathds{1}$, until a recurrence
occurs.

The second native TEDOPA parameter is the local dimension~$d$ of the
single oscillators constituting the environment. For a given
temperature, the occupation of the single oscillators can be determined
exactly, giving a rough scale of the necessary truncation level. Some
error will be introduced necessarily but this can be upper-bounded
analytically as explained in Ref.~\cite{ref:woods2015}. On the other
hand, numerical benchmark calculations with increasing local dimensions
will generally yield sufficiently accurate results.

A further subtlety in the chain mapping consists in the determination
of an adequate hard cutoff frequency~$ω_{hc}$ of the spectral density. For instance,
the slow approach to zero for large frequencies of the Drude-Lorentz
bath imposes a careful convergence check of the resulting physical
behaviour. For further discussions of these effects refer to
\cite{ref:woods2014}.

While some error sources (like the cutoff in the chain length)
introduce, if treated correctly, virtually no error at all, the matrix
size~$χ$ necessarily does so due to the nature of the MPO. However, as
already studied in the context of the TEBD algorithm, this error can be
monitored during the time evolution \cite{ref:schollwoeck2011}.
This results in a quantity very similar to the {\em discarded weight}
known from DMRG
\begin{equation}
	w_{\text{discarded}} = 1 - \sum_i e²_i.
	\label{eq:def-discarded-weight}
\end{equation}
This quantifies the deviation from the targeted state using the
discarded eigenvalues $e_i$. This error propagates in a non-trivial fashion and it is advisable to perform convergence checks in the dynamics under variation of the size of $χ$. An additional source of error is derivated from the Suzuki-Trotter decomposition used in
the TEBD part of TEDOPA.

It should be noted that the magnitude of the singular values
kept during MPO-procedures should not fall below some threshold $e_0$.
The transfer tensors determined by TTM do decay rapidly, falling to
comparatively low magnitudes, and singular values corrupted by
numerical noise deteriorates the interpretation of
results as well as the propagation procedure.

Finally, for TTM the important quantity to keep track of in simulations
is the norm of the memory kernel. This corresponds to the norm of the
transfer tensors divided by the squared time step~$δt$. This magnitude
should exhibit a sufficiently fast decay so that the remaining tail can
be neglected. Additionally, the time step $\delta t$ must be such that
it provides a good resolution of the features of the memory kernel.

\section{Benchmark \label{sec:Benchmark}}

In this section we verify the combination of TEDOPA with TTM by
comparing the obtained transfer tensors with those originating from
another numerically exact simulation method for non-Markovian systems
under the same conditions. The chosen benchmark regime is the Ohmic
Drude-Lorentz bath and the additional simulation method is the
hierarchy of equations of motion (HEOM) \cite{ref:tanimura1989}.

We consider the spin-boson model (SBM) and define the (monomeric) system Hamiltonian
\begin{equation}
 H_{\text{sys}} = \frac{1}{2}ε σ_z + \frac{1}{2}Δ σ_x.
\end{equation}
Here we set $\hbar=1$, a convention we will stick to from now on, and express all frequencies in units of $\epsilon$. We employ the standard SBM notation where $ε$ corresponds to the
energy bias between ground and excited state, $Δ$ is the tunneling
matrix element, and $σ_i$ $(i=x,y,z)$ are the Pauli matrices corresponding
to the $i$'th spatial direction. The system interaction operator $A$ is
defined as the excited state projector $|e\rangle\langle e|$. We
choose the parameter $Δ=0.6\epsilon$, and an Ohmic spectral density
of the Drude-Lorentz form
\begin{equation}
	J\left( ω \right) = λ~γ~\frac{ω}{\left( ω² + γ² \right)},
	\label{eq:jdrude}
\end{equation}
with parameters $λ=\epsilon$ and $γ=10\epsilon$ respectively identifying a scaling
of the interaction strength and a soft cutoff frequency. Thus the bath
reorganization energy $λ_r=\int J\left( ω \right) d ω $ is $λ_r=5.89\epsilon$. A
large hard cutoff $ω_{\text{hc}}=320\epsilon$ has been employed to meet the
aforementioned convergence requirements of TEDOPA under Drude-Lorentz
baths.

At an inverse temperature of $β=0.5\epsilon$, TEDOPA exhibits favorable cutoffs
and HEOM simulations are accurate, which enables benchmarking. The
resulting elements of the memory kernel obtained by TTM applied to
TEDOPA's initial trajectories are compared with those retrieved from
HEOM \cite{ref:tanimura1989} simulations of the same system. We have
confirmed agreement in a broad range of additional regimes accessible
to both TEDOPA and HEOM. Further the system's Hamiltonian has
successfully been recovered from the first transfer tensor. This also
corroborates the ability of TTM to extract the same dynamical tensors
irrespective of the simulation method used for the generation of the
trajectories. We will now turn to applications on hitherto inaccessible
regimes to illustrate the strengths of the TEDOPA-TTM combination.

\section{Applications \label{sec:Applications}}

\subsection{Non-Ohmic spectra}

\begin{figure}[h]
	\centering
	\includegraphics[width=0.9\columnwidth]{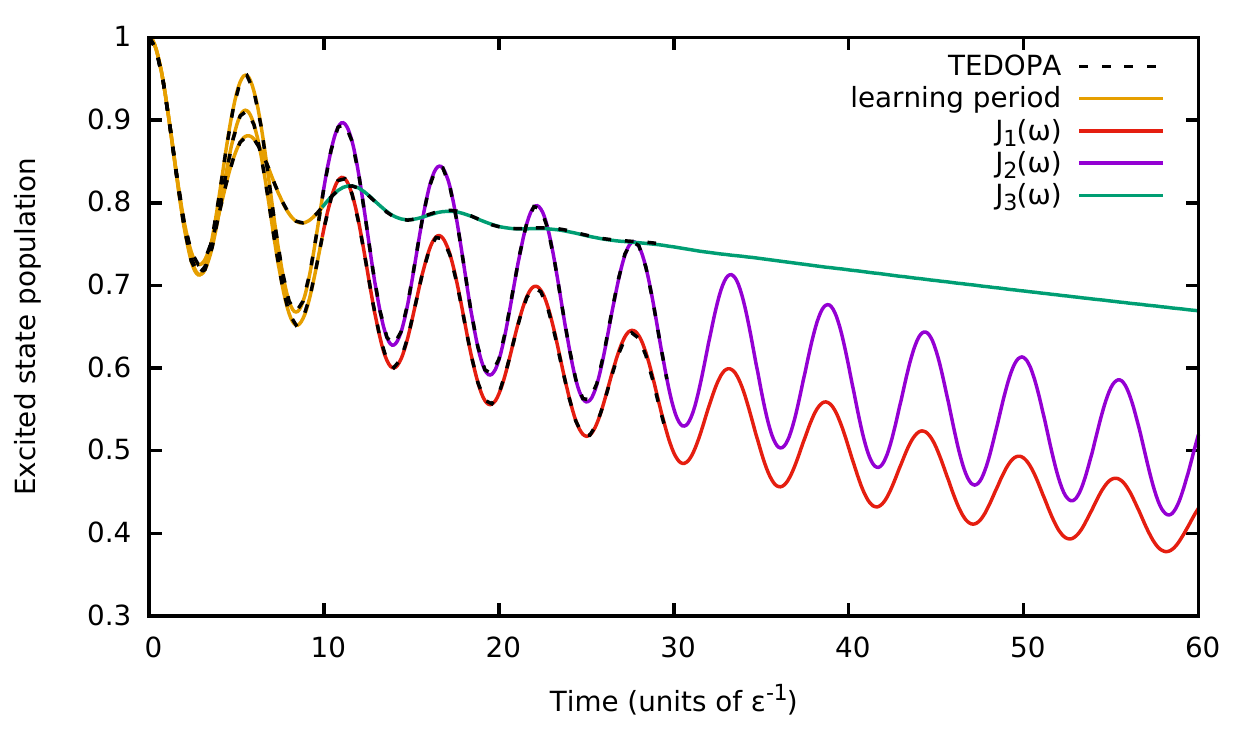}

	\caption{Effect on the population dynamics of the spin of three different non-Ohmic spectral
		densities $J_{1,2,3}$ (see main text for functional forms) at inverse temperature $β=\epsilon$
		and verification of the predictability of trajectories
		by TTM. Black dashed lines are TEDOPA simulation results,
		colored lines are TTM's predictions; TTM learning times
		are denoted by orange lines (roughly until $t=10\epsilon^{-1}$).}

	\label{fig:diff_J}
\end{figure}
By construction, TEDOPA is inherently suited to treat spectral
densities of arbitrary shape. When considering non-Ohmic spectral densities,
Markovian approaches are well-known to anomalously suppress the effect of pure dephasing contributions \cite{ref:frank2008, ref:chin2012}. In this
section we present an analysis of the dynamical effects of three
different non-Ohmic spectral densities, namely
\begin{align}
	J_1\left( ω \right) &= λ_1~ω^3~e^{-ω/ω_c}, \\
	J_2\left( ω \right) &= λ_2~ω^5~e^{-ω/ω_c}, \\
	J_3\left( ω \right) &= λ_3~√ω ~e^{-ω/ω_c}.
	\label{eq:diff_J}
\end{align}
To facilitate comparison, all of them exhibit the same exponential
decay with $ω_c=0.3\epsilon$ and are subject to a hard cutoff at $ω_{hc}=10\epsilon$.
Also the parameters $λ_{1}=1.8\epsilon$, $λ_{2}=1.0\epsilon$ and $λ_{3}=0.6\epsilon$ have been
chosen in such a way that they all share the same reorganization energy $λ_{r}=0.3\epsilon$. Thus the average interaction
strength between system and environment is the same and the functional
form of the spectral density is the factor responsible for disparate
dynamics.
The resulting dissimilar amplitudes and decay rates of the
oscillations due to the different spectral densities are illustrated in
Fig.~\ref{fig:diff_J}. The tunneling strength is, as in previous section, $Δ=0.6\epsilon$. One can observe that, for the fastest bath $J_2$, oscillations are sustained for a longer time, while this ability decreases for spectral densities centered in lower frequencies $J_1$ and almost disappears for the very slow bath represented by $J_3$. Some brief initial time is sufficient to generate the transfer tensors and predict the further evolution, whereupon high-accuracy TEDOPA
simulations are used to verify these predictions.

\begin{figure}[h]
	\centering
	\includegraphics[width=0.9\columnwidth]{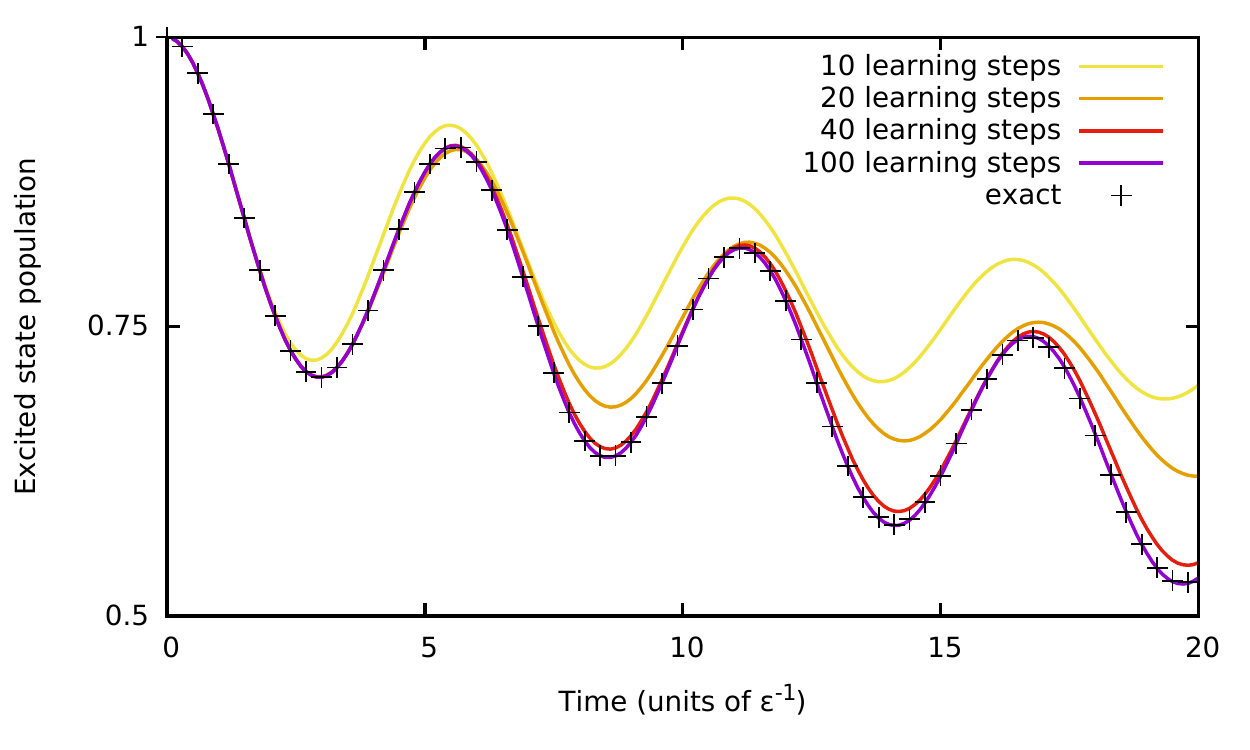}

	\caption{Time evolution of the excited state population subject
		to an environment with super-Ohmic spectral density $J_2\left( ω
		\right)$ at $β=\epsilon$.  Black crosses denote the TEDOPA-only
		evolution, while the TTM predictions (colored lines)
		show a gradual convergence upon increased learning time. The full $100$ learning steps correspond to time
		$t=10\epsilon^{-1}$.}

	\label{fig:learning}
\end{figure}
  The suitability of the TEDOPA-TTM combination
is supported by the fact that these simulations require on the order of
just 100 tensors to converge to the exact results that have been
obtained by full TEDOPA simulations, as shown in
Fig.~\ref{fig:learning}. This translates into about an order of magnitude faster results for TTM-TEDOPA combination than for TEDOPA alone. Further improvements in simulation speed are possible and are discussed in section \ref{sec:SimTime}.

\subsection{Low temperatures}

\begin{figure}[h]
	\centering
	\includegraphics[width=0.9\columnwidth]{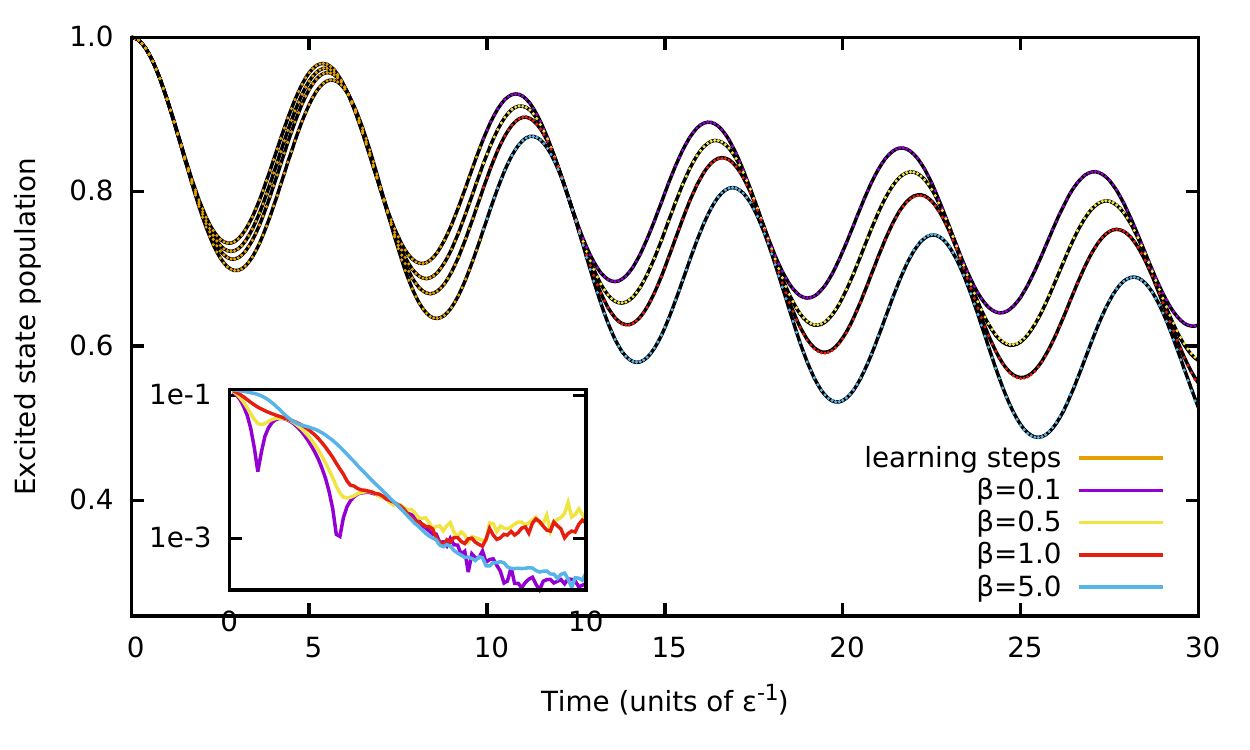}

	\caption{Decay of a monomeric system's population in the SBM, subject to
		an environment with super-Ohmic spectral density $J_2\left( ω
		\right)$ at different inverse temperatures~$β$. For
		better clarity only the first few oscillations
		are shown. Solid lines correspond to TEDOPA-only
		simulations with verified accuracy. The orange initial
		part of each curve corresponds to the learning period.
		The decay of the tensor norm for the learning period is
		shown in the inset.}

	\label{fig:Tseries_x5}
\end{figure}

\begin{figure}[h]
	\centering
	\includegraphics[width=0.9\columnwidth]{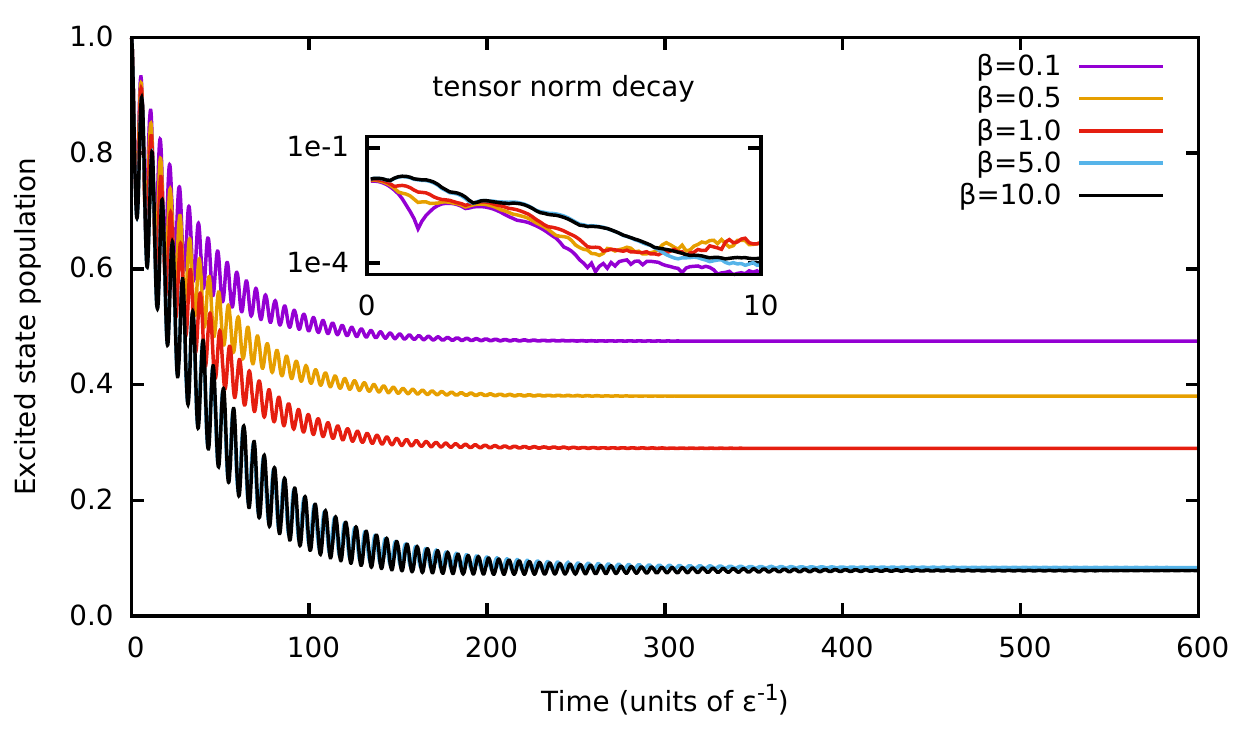}

	\caption{Thermalization of the system's population subject to
		an environment with spectral density $J_1\left( ω
		\right)$ at different temperatures. The combination TEDOPA-TTM has been used and verified with TEDOPA-only simulations. The decay of the
		tensor norm for the learning period is shown in the
		inset.}

	\label{fig:Tseries_x3}
\end{figure}

\begin{figure}[h]
	\centering
	\includegraphics[width=0.9\columnwidth]{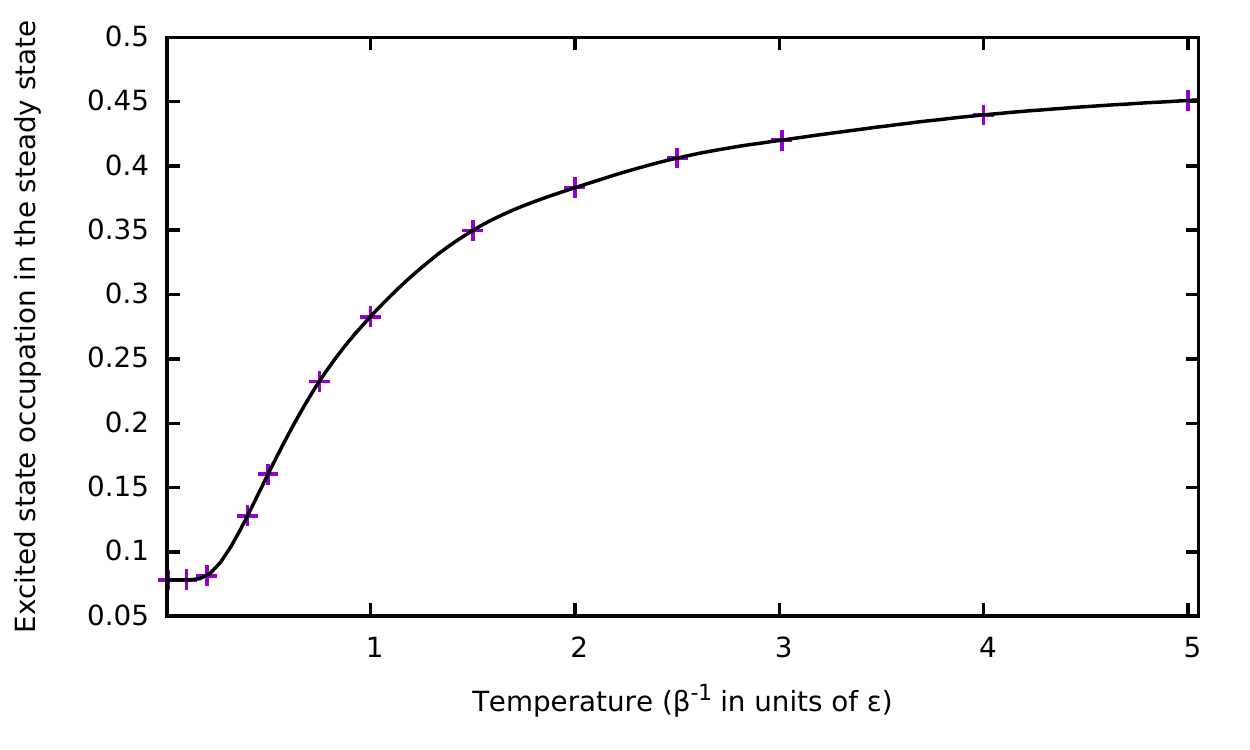}

	\caption{Excited state population in the steady state for a
		monomeric system subject to spectral density
		$J_{1}\left( ω \right)$, plotted over the inverse
		temperature~$β$. The steady state is determined by TTM
		evolution of the initial TEDOPA trajectories. The line is a guide to the eye.}

	\label{fig:steadystate}
\end{figure}

To further illustrate the power of our approach, we present results for
a broad range of low to very low temperatures, up to $β=10\epsilon$. For the super-Ohmic
spectral density $J_2\left( ω \right)$ we show in Fig.~\ref{fig:Tseries_x5} that it is possible to
simulate the dynamics of a monomeric system at various inverse
temperatures and the same
system parameters as in the previous example. For the case of spectral
density $J_1\left( ω \right)$ we employ TTM to propagate the system
until the steady state is reached (Fig.~\ref{fig:Tseries_x3}) and plot
the steady-state occupation of the excited state for various inverse
temperatures~$β$ in Fig.~\ref{fig:steadystate}. The insets in
Figs.~\ref{fig:Tseries_x5} and \ref{fig:Tseries_x3} show the memory
kernel's decay over several orders of magnitude for the corresponding
spectral densities. It is this decay which certifies the possibility to
use the tensors for long-time propagation of the dynamics.

\subsection{Absorption spectrum}

The combination of TEDOPA and TTM is especially indicated for
applications where accurate simulation of long time dynamics is
crucial. The determination of absorption spectra belongs to this class
of problems and we analyse here the more complex case of a dimeric
system consisting of two coupled monomers.

The coupled dimeric system in the single excitation manifold consists of two excited states $| e_1 \rangle $, $| e_2 \rangle $ and a common ground state $| g \rangle $, and is described by the Hamiltonian
\begin{equation}
	H_{sys} = ε_1 |e_1\rangle\langle e_1| + ε_2 |e_2\rangle\langle e_2| + J
	\left( |e_1\rangle\langle e_2| + |e_2 \rangle\langle e_1| \right),
	\label{eq:dimer}
\end{equation}
where parameters $ε_1\equiv\epsilon$, $ε_2=2\epsilon$, and exchange interaction strength
$J=0.6\epsilon$ are chosen. Each of the two systems is coupled to a bath. Without loss of
generality we assume both environments are described by the same
spectral density $J_1\left( ω \right)$ and at temperature $β=\epsilon$.

The absorption spectrum is calculated as the Fourier transform of the
two point correlation function of the dipole operator
$\hat{μ} = \mu_1 |e_1\rangle\langle g| + \mu_2 |e_2\rangle\langle g|+ h.c.$
\begin{align}
	C_{μ-μ} \left(t, 0\right) &=
	\left\langle \hat{μ}\left(t\right)
	\hat{μ}\left(0\right)\right\rangle \\
	&= \text{tr}\left[ e^{-iHt} \hat{μ} e^{iHt} \hat{μ}
	ρ\left( 0 \right) \right],
\end{align}
between times $t=0$ and $t=τ$ such that the steady state has been reached at $\tau$.

\begin{figure}[h]
	\centering
	\includegraphics[width=0.9\columnwidth]{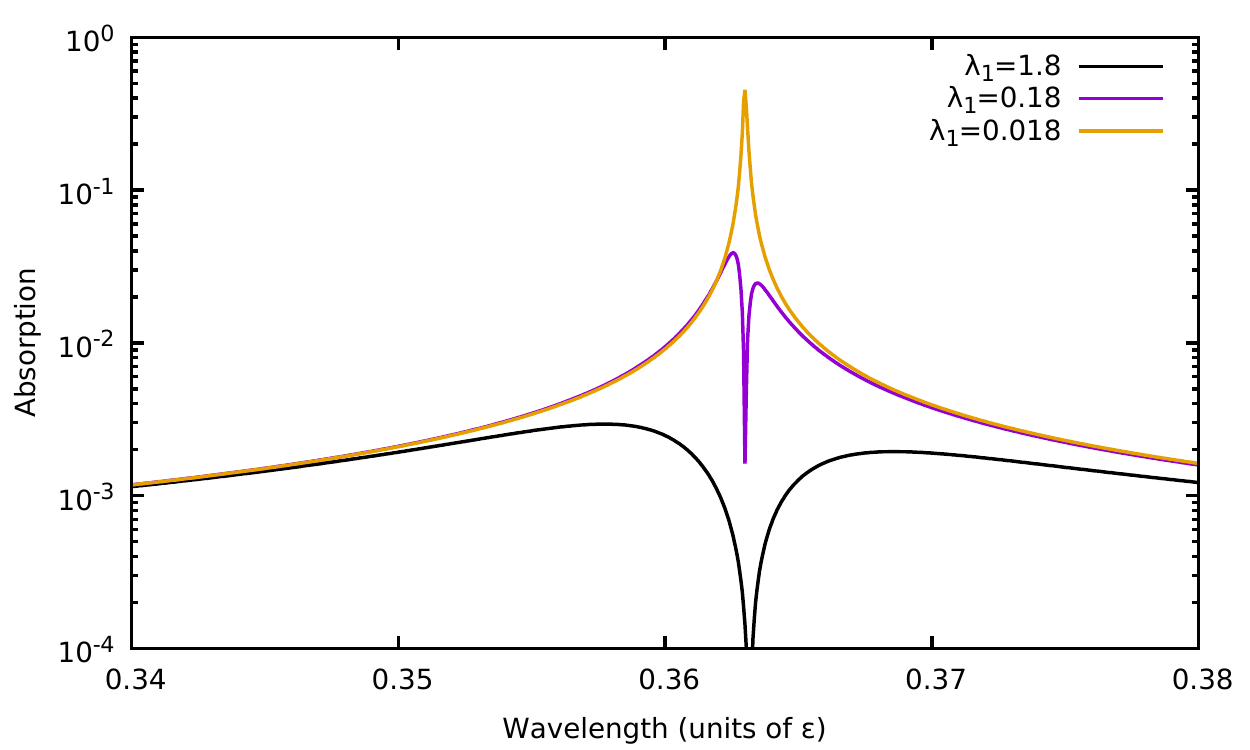}

	\caption{Peak structure of the absorption spectrum of a dimeric system for different values of the system-environment coupling strength. The emergence of the vibrational fine structure is apparent for increasing strength of the coupling to the environment.}

	\label{fig:spectrum}
\end{figure}

In the limit of weak interaction with the environment, the absorption
spectrum emerging from Hamiltonian Eq.\eqref{eq:dimer} exhibits two
peaks in the one-exciton subspace. One of them is shown in the $\lambda_1=0.018\epsilon$ line (green) of Figure~\ref{fig:spectrum}, corresponding to the second excited state in the excitonic manifold at a wavelength of around $0.363\frac{c}{\epsilon}$. For higher coupling strengths with the environment, the emergence of the vibrational fine structure splits the peak in two, which is shown in the $\lambda_1=0.18\epsilon$ line (blue) and the $\lambda_1=1.8\epsilon$ line (black).

It will be interesting to compare the efficiency of the approach presented
here with 
other approaches such as stochastic path integral methods which has recently
been 
developed to calculate absorption and emission spectra [17] specifically for
low temperatures 
and long times.

\subsection{Simulation time \label{sec:SimTime}}

The ability of the TEDOPA-TTM
combination to explore new simulation regimes is a consequence of
the extraordinary savings in computational resources. We will explore
these in terms of the ``wall time'' $t_w$, the physical time required for the
simulation to be executed as measured by an external clock.

Three factors have a direct influence on simulation time:
\begin{itemize}
	\item bath coherence time $t_{\text{bath}}$,
	\item chain length $N$ and
	\item system dimension $d_{\text{sys}}$.
\end{itemize}
TTM requires the simulation of $d_{\text{sys}}^2$ trajectories until $t_{\text{bath}}$, one for each independent
initial density matrix. Although this overhead may become inconvenient
for systems of large dimension, the computation may be parallelized to avoid a scaling of $t_w$ with $d_{\text{sys}}$. Even without parallelization, numerical studies often require exploration of a large number of independent initial conditions anyway.

Due to the efficiency of multiplicative propagation with TTM (Eq.\ref{eq:TTM}), nearly the totality of the wall time $t_w$ required for a simulation until $t_{\text{sim}}$ is employed in the initial generation of the tensors until $t_{\text{bath}}$ with TEDOPA. Therefore, one may consider $t_w$ to be essentially independent of $t_{\text{sim}}$. This makes the TTM-TEDOPA combination suitable for long time simulations, i.e.~cases where $t_{\text{sim}}\gg t_{\text{bath}}$. There is an additional benefit in shortening simulations with TEDOPA to $t_{\text{bath}}$, since this reduces the necessary chain length $N$.

\begin{figure}[h]
	\centering
	\includegraphics[width=0.9\columnwidth]{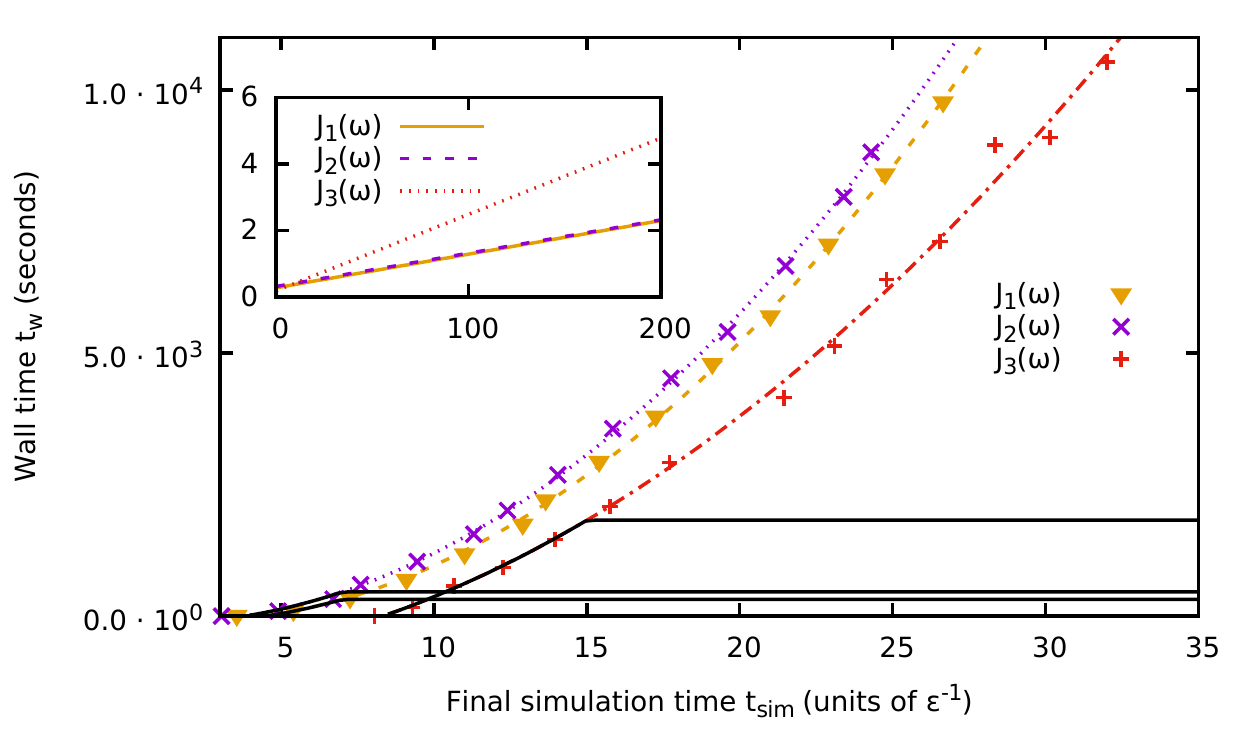}

	\caption{Data points show single-core TEDOPA wall times $t_w$
		for a specific simulation time $t_{\text{sim}}$; the
		respective line of the same color is the corresponding
		quadratic fit. Black lines show the simulation time for
		the same physical setting upon employing the
		combination of TEDOPA and TTM. The TTM-part grows
		linearly as can be seen in the inset (on the main panel
		the slope of these lines is so small that they appear horizontal). Note the different scales on the vertical axis between main plot and inset.}

	\label{fig:t_simulation}
\end{figure}
The scaling of the wall time $t_{w}$ necessary to perform a TEDOPA
simulation of timestep $δt$ until $t_{\text{bath}}$ can be expressed as
\begin{equation}
	t_w \propto N \frac{t_{\text{bath}}}{δt} \bar t,
	\label{eq:def-tw}
\end{equation}
where dependence on three factors has been made explicit: the number of sites $N$, the number of
time steps $\frac{t_{\text{bath}}}{δt}$ and a factor~$\bar t$ denoting the average wall time
necessary to simulate one chain site during one time step~$δt$.
However, in order to avoid end-of-chain recurrences, for a simulation time~$t_{\text{bath}}$ one requires
\begin{equation}
	N \propto t_{\text{bath}} \cdot \bar v,
	\label{eq:def-N}
\end{equation}
sites~$N$ in the environment, given an average propagation speed $\bar v$
in the chain. Thus a total wall time of
\begin{equation}
	t_w \propto \frac{\bar v \bar t}{δt}  t_{\text{bath}}^2 \equiv c
	\cdot t_{\text{bath}}^2,
	\label{eq:def-tw2}
\end{equation}
is needed where $c$ is a scenario-dependent constant.

The global speedup provided by the TTM-TEDOPA combination is illustrated in Fig.~\ref{fig:t_simulation}
for three instances with different spectral densities. The near independence of $t_w$ on $t_{\text{sim}}$ is shown for large $t_{\text{sim}}$. As shown in the inset, in reality $t_w$ increases linearly with $t_{\text{sim}}$, although with a negligible slope. For simulations with TEDOPA alone, the quadratic dependence expressed in Eq.(\ref{eq:def-tw2}) extends beyond $t_{\text{bath}}$ until $t_{\text{sim}}$.

\section{Conclusion and Outlook}

In this work we demonstrated that the combination of TEDOPA and TTM result in an enhanced simulation method of general non-Markovian open-quantum-systems especially well-suited for (but
not restricted to) low-temperature regimes and highly structured spectral densities. The formulation in terms of a multiplicative operator whose size is independent of the goal simulation time facilitates exploration of much longer, so far inaccessible timescales.

We verified the feasibility of this combination by a benchmark and presented applications for various spectral densities to highlight the
flexibility of our method. Further to the paradigmatic examples
presented, even larger benefits can be expected
upon application to simulations which are post-processed by some
averaging-type method. These are often noise-tolerant or noise-stable,
so small deviations do not change the characteristic features of the
final result. This type of analysis are expected to be of crucial importance for providing accurate microscopic models of the dynamical behaviour of mesoscopic systems and therefore a better understanding of how coherent effects still manifest in those time and length scales \cite{ref:Huelga2013,ref:levi2015,ref:rienk2011,Mohseni2014}.




\section{Acknowledgments}

This work was supported by the Alexander von Humboldt-Professorship, the
EU Integrating project SIQS, the EU STREP projects PAPETS, QUCHIP and EQUAM,
National Science Foundation (NSF) (Grant No. CHE-1112825) and Defense Advanced Research Projects Agency (DARPA) (Grant No. N99001-10- 1-4063), the MIT-Germany Seed Fund, and
the ERC Synergy grant BioQ. Computational resources used included bwUniCluster,
supported by the Ministry of Science, Research and the Arts
Baden-W\"{u}rttemberg and the Universities of the State of
Baden-W\"{u}rttemberg, Germany, within the framework program bwHPC.


\bibliographystyle{apsrev4-1}

\bibliography{transfer_tensors}{}

\end{document}